\documentstyle{article}
\begin{document}
\begin{titlepage}
\begin{centering}
\title{Scale Dependence of Polarized DIS Asymmetries\thanks{Partially
supported by CONICET-Argentina.}}
\author{D. de Florian$^{1}$, C.A.Garc\'{\i}a Canal$^{1}$,  S.Joffily$^{2}$,
R.Sassot$^{3}$ \\ \\
$^{1}$Laboratorio de F\'{\i}sica Te\'{o}rica \\
Departamento de F\'{\i}sica \\
Universidad Nacional de La Plata \\ C.C. 67 - 1900 La Plata \\
Argentina \\ \\
$^{2}$ Centro Brasilero de Pesquisas Fisicas \\ Rua Xavier Sigaud 150, Urca,
\\22.290.180 Rio de Janeiro \\
Brazil\\ \\
$^{3}$Departamento de F\'{\i}sica \\
Universidad de Buenos Aires \\
Ciudad Universitaria, Pab.1 \\
(1428) Bs.As. \\
Argentina}
\date{18 May 1995}
\maketitle
\end{centering}

\begin{abstract}
We compare the $Q^{2}$ dependence of the polarized deep inelastic scattering
proton asymmetry, driven by the leading order Altarelli Parisi evolution
equations, to those arising from fixed order $\alpha_{s}$ and $\alpha_{s}^{2}$
approximations. It is shown that the evolution effects associated with gluons,
which are not properly taken into account by the leading order approximation,
cannot be neglected in the analysis of the most recent experimental data.
\end{abstract}

\end{titlepage}

\noindent{\large \bf Introduction:}\\

Within the last few years, several analysis have been made on the scale
dependence of the polarized deep inelastic scattering structure
function $g_{1}(x,Q^{2})$, and the related asymmetry $A_{1}(x,Q^{2})$
\cite{Forte,Gehr,Lead,Altn},
which has been measured by different experimental groups
\cite{Slac,EMC,SMC,E143}.
Accurate estimates of the magnitude of the scaling violations in these
quantities
are essential ingredients in the understanding and interpretation of the
experimental data. These data are taken at different values of the scale and
then
used to determine other quantities defined at a common scale, such as the
moment of the structure function or parametrizations of parton distributions.
The increase in the precision of the experimental data, and the extension of
the kinematical range attained in the latest measurements
\cite{SMC,E143},
calls for a careful discussion of the different approximations implemented in
order to estimate the scale dependence of the data and its consequences in the
interpretation of the experiments.

{}From the theoretical point of view, the main obstacle in the
study of the scale dependence comes from the combination of two factors: while
the gluon contribution to the structure function, which may be large and
essential
in the partonic interpretation of the experiments, enters at next-to-leading
order (NLO) of perturbative QCD, the corresponding evolution equations have
not been calculated yet. In face of this, in most of the attempts, the strategy
consists in using the well known leading order (LO) Aaltarelli Parisi (AP)
evolution kernels, with
quark and gluon distributions defined either at NLO
\cite{Forte},
or LO but with an ad hoc gluonic term
\cite{Gehr,Lead,Altn}.

While there exists a complete freedom in the choice for the definition of the
parton distributions, provided the choice is implemented consistently in other
processes, it is clearly inconsistent to evolve them with evolution equations
obtained in other schemes of definition. Moreover, a large gluon contribution,
as
suggested in many analysis of the experimental data
\cite{plspin,Lead},
may have a crucial role in the evolution of the
structure function and it is not clear a priori wherever the AP LO kernels,
calculated at an order where there is no gluon contribution to the structure
function, will properly account for its role.

Fortunately, there exists an alternative to the usual AP evolution method
which may bypass the obstacle mentioned previously. This alternative is based
in what is called fixed order perturbation theory and was presented in
references
\cite{VanNg,VanNf}
in connection with the problem of the evolution of $g_{1}(x,Q^{2})$. Within
this method it is possible to write down the structure function in terms of
parton distributions at order $\alpha_{s}$ or $\alpha_{s}^{2}$ and evolve them
consistently, exploiting a convenient choice of the factorization scale, which
shifts the scale dependence from the parton distributions to already known
coefficients. Both fixed order calculations approximate the AP results at
LO and NLO, resumming one or two powers of $log(Q^{2}/M^{2})$ respectively.

In this paper we implement the above mentioned method in the analysis of the
polarized asymmetries using  well defined sets of parton distributions. One of
them with a large gluon density and another without it. Both sets are designed
in order to reproduce the asymmetry within the present experimental errors.
First, we verify that the  fixed order $\alpha_{s}$ calculation approximates
the LO AP results in a wide range of the scale. Then we show that the
 fixed order $\alpha_{s}^{2}$ evolution, almost equivalent to the NLO AP
result,
differs from the available AP calculations for the set with a large gluon
component and discuss the reason. Finally,
we calculate the effects  of the correct evolution in the data and the
moments of the structure functions and compare these results with those
obtained via AP like approximations.\\

\noindent{\large \bf Definitions}\\

In order to unambiguosly define what we mean by fixed order perturbation at a
given
order, LO and NLO evolution, and our specific choice  for the definition of
parton distributions,
we begin by writing in equation (1) the general expression for the structure
function
$g_{1}(x,Q^{2})$ in terms of parton distributions, as given in the improved
parton model
\cite{VanNg}
\begin{eqnarray}
g_{1}(x,Q^{2}) &= &\frac{1}{2} \int_{x}^{1} \frac{dz}{z}
\left [ \frac{1}{n_{f}} \sum_{k=1}^{n_{f}} e_{k}^{2} \left \{\Delta q^{S}
\left (\frac{x}{z},M^{2} \right ) C_{q}^{S} \left ( z,\frac{Q^{2}}{M^{2}}\right
) + \right. \right.\\
& & \left. \left. \Delta g\left (\frac{x}{z},M^{2}\right )
C_{g}\left ( z,\frac{Q^{2}}{M^{2}}\right )\right \} +   \Delta q^{NS}\left
(\frac{x}{z},M^{2} \right ) C_{q}^{NS}\left ( z,
\frac{Q^{2}}{M^{2}}\right ) \right ] \nonumber
\end{eqnarray}
 $\Delta q^{S}$ denotes the singlet combination
of the polarized quark ($\Delta
q_{k}$) and antiquark ($\Delta \overline{q_{k}}$) densities
 of species $k$,
\begin{equation}
\Delta q^{S}(z,M^{2})=\sum_{i=1}^{n_{f}} [ \Delta q_{i}(z,M^{2}) + \Delta
\overline{q_{i}}(z,M^{2}) ]
\end{equation}
whereas $\Delta q^{NS}(z,M^{2})$ denotes the nonsinglet combination,
\begin{equation}
\Delta q^{NS}(z,M^{2})= \sum_{i=1}^{n_{f}}\left (e_{i}^{2}-\frac{1}{n_{f}}
\sum_{k=1}^{n_{f}} e_{k}^{2} \right )
[ \Delta q_{i}(z,M^{2}) + \Delta \overline{q_{i}}(z,M^{2}) ]
\end{equation}
and $\Delta g$, the polarized gluon density. The coefficient functions
$C_{q,g}$
 that
multiply each combination of parton distributions are labeled correspondingly
and can be calculated at a given order in $\alpha_{s}$ once the prescriptions
for the regularization of ultraviolet singularities and the factorization of
those infrared of collinear origin are adopted.
Consequently, parton distributions introduced in this way depend  on the
order of perturbation,  on the ultraviolet regularization,  and on  the
factorization procedure.

At order $\alpha_{s}^{0}$, the singlet and non singlet coefficients reduce to
the $\delta (1-z)$ function and the gluon coefficient vanishes. There is no
need to specify any prescription due to the absence of singularities. At order
$\alpha_{s}^{1}$, there are two classes of contributions to the coefficient,
one whose dependence in $Q^{2}/M^{2}$ is logarithmic and another which is not.
At order $\alpha_{s}^{2}$ the contributions are classified according to they
have $\alpha_{s}^{2}log^{2}(Q^{2}/M^{2})$, $\alpha_{s}^{2}log(Q^{2}/M^{2})$,
$\alpha_{s}log(Q^{2}/M^{2})$ or non logarithmic dependence.
Both the $\alpha_{s}^{2}$ and the $\alpha_{s}$ have been calculated in
references
\cite{VanNg}
for different factorization prescriptions. There has been a long debate about
the
way in which the factorization of collinear contributions can be made
\cite{Vogel}.
We adopt the procedure described in reference
\cite{prsw}
which leaves opened the possibility of a non vanishing gluon contribution to
the structure function.

The scale $M$ in equation (1), often called factorization scale, is a relic of
the factorization procedure and indicates the scale that separates partonic
from hadronic effects in the definition.  This scale, which in principle would
be arbitrary, i.e. provided the coefficient functions were calculated up to
infinite order, allows two strategies for the study of the scale dependence
of the structure functions. The most used consists in redefining parton
distributions in such a way they absorb the scale dependence of the
coefficients, and then choosing $Q^{2}=M^{2}$. In this way the dependence on
the physical scale $Q^{2}$ is shifted from the coefficients to the parton
distributions.
Then one can use the scale analogously to the renormalization scale, in
the procedure that removes the ultraviolet divergences leading to the running
coupling constant. A similar  procedure in this case leads to the familiar AP
equations
\cite{APrep}.

What is called LO evolution amounts to calculate up to order $\alpha_{s}$
the coefficient functions in equation (1) and absorb only the
$\alpha_{s}logQ^{2}$
term in the redefinition of the parton distributions, thus neglecting the non
logarithmic terms. As the gluonic contribution to the structure function enters
through non logarithmic terms, it is not present in this approach. The use of
the renormalization group techniques whithin the AP evolution equations makes
them take into account effectively not only contributions like
$\alpha_{s}logQ^{2}$,
characteristic of the order $\alpha_{s}$ diagrams, but the whole series of
powers of $\alpha_{s}logQ^{2}$. The NLO approximation, in turn, takes into
account contributions like $\alpha_{s}^{2}logQ^{2}$ and absorbes all the
$\alpha_{s}$ terms in the redefinition of parton distributions. Then, gluons
contribute to the structure functions at NLO, however the coefficients that
drive their scale dependence have not been calculated yet.

The second strategy amounts to keep $M^{2}$ fixed and let the $Q^{2}$ evolution
of the structure function proceed via the $log(Q^{2}/M^{2})$ terms in the
coefficients.  This is called fixed order perturbation theory and leads to the
same results of the first method provided $\alpha_{s}log(Q^{2}/M^{2})<<1$.
Satisfied this condition, the leading order evolution is approximated by the
fixed order $\alpha_{s}$ method, as those higher powers of
$\alpha_{s}log(Q^{2}/M^{2})$,
present only in AP LO, are not significant. By the same reason, NLO is almost
equivalent to fixed order $\alpha_{s}^{2}$; in both approaches terms like
$\alpha_{s}log(Q^{2}/M^{2})$, $\alpha_{s}^{2} log^{2}(Q^{2}/M^{2})$ and
$\alpha_{s}^{2}log(Q^{2}/M^{2})$ are taken into account, the difference being
in terms like  $\alpha_{s}^{3}log^{3}(Q^{2}/M^{2})$,
$\alpha_{s}^{3}log^{2}Q^{2}/M^{2})$
and higher. The same argument applies for the unpolarized structure functions
and in that case this has been verified
\cite{VanNf}.
As the NLO evolution is not feasible in polarized case, it seems sensible to
use fixed order $\alpha_{s}^{2}$ evolution.\\

\noindent{\large \bf Polarized parton distributions}\\

In order to evaluate the actual scale dependence of the structure functions,
and through them of the asymmetries, in both evolution strategies one needs a
set of parton distributions, at a given scale and defined in accordance to the
coefficients or kernels to be used. The problem one confronts then is that the
data points coming from the available experiments are given for different
values of $x$ and $Q^{2}$.  In the unpolarized case
\cite{Stir}
the problem is solved
assuming certain functional $x$-dependence for the parton distributions and
adjusting the parameters in the functional form in such a way the evolved
distributions reproduce the data in an iterative process.
The number of points to be fitted is more than 30 times greater than that of
the paramenters. In the polarized case however, not only the NLO evolution
kernels are missing, but the number of points is much more reduced. One has
then to look for additional constraints on the parton distributions  and
perform the analysis at LO. In reference
\cite{Gehr}
this approach has been followed with a slight variation, which is addmitedly
ambiguous, in order to include the gluon contribution to the structure
function.
In previous analysis to that, a coarser approximation was implemented assuming
the asymmetry to be essentially independent of the scale (for a comprehensive
review see reference
\cite{Lead}).

In the following we show explicitly the first steps of the iterative procedure.
This will allow us to estimate the size of the corrections the experimental
data for the asymmetries get when reduced to a common $Q^{2}$ value.

The first step in our program consists in obtaining a set of parton
distributions
that, at some  average energy scale, lead to asymmetry values compatible with
those obtained experimentaly. The asymmetry is taken to be given by
\begin{equation}
A_{1}(x,Q^{2})\simeq \frac{g_{1}(x,Q^{2})}{F_{1}(x,Q^{2})}
\end{equation}
where $g_{1}(x,Q^{2})$ is that of equation (1) using coefficients calculated
up to order $\alpha_{s}^{2}$ with the factorization prescription already
specified. Both $M^{2}$ and $Q^{2}$ are set to be equal to $10\,GeV^{2}$, which
is an average value of the  scales studied by the experiments and will
allow us to go to different values of $Q^{2}$ keeping
$\alpha_{s}log(Q^{2}/M^{2})$
small enough. The structure function $F_{1}(x,Q^{2})$ is taken to be given
by its $\alpha_{s}^{2}$ expression, fed with the very recent set of unpolarized
parton distributions of reference
\cite{Stir}.
The result can be seen in figures 1 and 2 for different sets of data and for
two sets of parton distributions, with a large amount of gluons ($\Delta
g=2.5\,\Delta s=0$) in set 1 and without gluons ($\Delta g=0, \Delta s=$)
in set 2. Both sets satisfy
also other constraints such as asymptotic behaviour and positivity of parton
distributions and those coming from hyperon $\beta$ decays \cite{prsw}

The second step estimates the error introduced by the assumption about an
almost scale independent asymmetry, comparing each data point with the evolved
asymmetry. As it can be appreciated in figures 3 and 4, whereas the set without
gluons induces a small evolution effect, the asymmetry calculated with parton
distributions with a large gluon component exhibits a significant scale
dependence. Although the evolution is not negligible, as we were forced to
assume in the first step, the actual $Q^{2}$ corrections calculated with set 1
(with gluons), rather than invalidate our set, improves the quality of the
fit. This improvement is particularly conspicuous when comparing SMC
\cite{SMC}
and E-143
\cite{E143}
low $x$ data to the asymmetry values calculated with the set.

Figures 5 an 6 show the asymmetry calculated at $10\,GeV^{2}$ compared to
the E-143 and SMC data taken at different values of $Q^{2}$ and rescaled to
$10\,GeV^{2}$. We rescale the experimental data calculating for each value
of $x$, the difference between the asymmetries at the measured scale and at
$10\,GeV^{2}$, both calculated using our sets as input.

The third and subsecuent steps, modify the original (previous) parametrization
in order to improve the quality of the fit. As the precision attained here is
more than we
need to illustrate our discussion, we stop here for the moment, and use our
two sets to compare the different evolution approaches.\\

\noindent{\large \bf Fixed order $\alpha_{s}$ and $\alpha_{s}^{2}$ evolution}\\

In this section we compare the evolution of the proton asymmetry as given by
the $\alpha_{s}$, $\alpha_{s}^{2}$ and LO approximations. In figure 7 we plot
the result of evolving the asymmetry from $10\,GeV^{2}$ (solid line) to
$3\,GeV^{2}$
using the three methods and set 1 ($\alpha_{s}$: long dashes, $\alpha_{s}^{2}$:
short dashes, LO: dots). For small values of $x$ there is a clear cut
difference
between the $\alpha_{s}^{2}$ and the other approximations. The origin of this
discrepancy can be associated
to the contributions which are proportional proportional to
$\alpha_{s}^{2}log(Q^{2}/M^{2})$ and to the
gluon distribution. Both of them are present in the $\alpha_{s}^{2}$
approximation
but not in the others.
In this region, the LO approximation produces similar results to those of the
$\alpha_{s}$. For values of $x$ greater than between  $0.1$, LO is better
approximated
by the $\alpha_{s}^{2}$, presumably due to the importance of greater terms
like
$\alpha_{s}^{2}log^{2}(Q^{2}/M^{2})$, which are included in the latter
approximations.
The same asymmetry, but calculated with set 2, do not show such a strong
dependence on the evolution method (figure 8).

The difference in the $Q^{2}$-dependence estimated by the approximations we
are analyzing, have non negligible consequences in the moment of the spin
dependent structure function $g_{1}$, which is calculated from the asymmetry
measurement. This can be seen in figure 9, where we show the moment as a
function
of the lower integration limit. This is computed rescaling the data with the
$\alpha_{s}^{2}$ and $\alpha_{s}$ approximations. The extrapolations were
estimated
integrating our sets of parton distributions in the unmeasured range. The
results are also given in Table 1.

Notice that the difference between the moments comes mainly from the low $x$
region, where the LO and the $\alpha_{s}$ approaches give similar results,
differing with the $\alpha_{s}^{2}$ estimate. In average the effect of the
$\alpha_{s}^{2}$ evolution is to reduce the value estimated with both sets,
as it was found in previous LO analysis, but with a more significative
correction.
The theoretical uncertainty introduced by the evolution is considerably
greater than previous analysis, as can be seen in Table 1.

Even though in average the correction to the moment is negative, for each
experiment the corrections show differents patterns. For example, analysing
E143 data we find them to be positive. This is due to the fact that these data
points lay in the region of $x$ where the asymmetries grows with $Q^{2}$, and
all of them have $Q^{2}$ values lower than $10\,GeV^{2}$. EMC data points
belong to the same region, however the evolution to $10\,GeV^{2}$ is in the
opposite direction thus giving  negative corrections. For SMC there is an
additional negative contribution coming from the low $x$ and low $Q^{2}$ data,
where the asymmetry decreases with the scale.
This emphasizes the importance of taking into account the
actual $x$ dependence of the scaling violations, which is not always done.
\\

\noindent{\large \bf Conclusions:}\\

The main conclusion of our analysis is that there exists a significant
difference between the $\alpha_{s}$ and $\alpha_{s}^{2}$ evolution of the
asymmetries at small values of the variable $x$. This means that a non
negligible
difference will be found between the LO and NLO analysis. As the small $x$
data, which is usually taken at low values of the $Q^{2}$-scale, are crucial in
the
estimation of the moments, these corrections must be taken into account.
The analysis beyond leading order emphasizes the difference between the
evolution at small and large $x$, which is essential when different sets of
data are compared.
Given that the fixed order technique is consistent and more accurate than the
ambiguous mixture of perturbation orders in the other approaches, it seems
sensible
to use it in forthcoming analysis of polarized deep inelastic scattering
data.\\

\noindent{\large \bf Acknowledgements:}\\

One of us (CAGC) acknowledges the Centro Brasileiro de Pesquisas Fisicas
(CBPF), Rio de Janeiro, Brazil for the warm hospitality extended to him
during a recent visit in the framework of the TWAS Associate Membership
Scheme at Centres of Excellence.

\pagebreak

\pagebreak
\noindent{\large \bf Figure captions}\\

\noindent{\bf Figure 1} Polarized deep inelastic scattering asymmetry data
coming from SMC and  E-143 experiments and calculated with set 1 and set 2 at
$Q^{2}=\,10GeV^{2}$.

\noindent{\bf Figure 2} Polarized deep inelastic scattering asymmetry data
coming from E-130, E-80 and EMC experiments and calculated with set 1 and set 2
at $Q^{2}=\,10GeV^{2}$.

\noindent{\bf Figure 3} $Q^{2}$ dependence of the asymmetry (set 1,
$\cal{O}$$(\alpha_{s}^{2})$)

\noindent{\bf Figure 4} $Q^{2}$ dependence of the asymmetry (set 2,
$\cal{O}$$(\alpha_{s}^{2})$)

\noindent{\bf Figure 5} The asymmetry at $Q^{2}=\,10GeV^{2}$ and the SMC data
rescaled to that value.

\noindent{\bf Figure 6} The asymmetry at $Q^{2}=\,10GeV^{2}$ and the E-143 data
rescaled to that value.

\noindent{\bf Figure 7} Evolution effects on the asymmetry coming from the
different methods (set 1).

\noindent{\bf Figure 8} Evolution effects on the asymmetry coming from the
different methods (set 2).

\noindent{\bf Figure 9} Evolution effects on the moments of $g_{1}^{p}$ coming
from
different methods and parton distributions.\\

\noindent{\large \bf Table caption}\\

\noindent{\bf Table 1} Moments of $g_{1}^{p}$ estimated assuming no scale
dependence in the asymmetry (naive), and $\alpha_{s}^{2}$ evolution (the
unmeasured interval calculated with the sets).

\vspace*{20mm}

\begin{center}
\begin{tabular}{|c|c|c|c|}\hline
{\small Exp.} &{\small Naive}&{\small Set 1 $\alpha_{s}^{2}$}&{\small Set 2
$\alpha_{s}^{2}$} \\ \hline
EMC   &   0.128 & 0.120 & 0.126 \\
E143  &   0.126 & 0.139 & 0.129 \\
SMC   &   0.134 & 0.113 & 0.130 \\ \hline
\end{tabular}
\\\vspace*{10mm}{\large \bf Table 1.}\\
\end{center}

\end{document}